\documentclass[journal]{IEEEtran}

\ifCLASSINFOpdf
\else
   \usepackage[dvips]{graphicx}
\fi
\usepackage{url}
\usepackage{graphicx}
\usepackage{multirow}
\usepackage{threeparttable}
\usepackage{amsmath}
\usepackage{amssymb}
\usepackage{cite,color}

\begin{document}

\title{Eeg2vec: Self-Supervised Electroencephalographic Representation Learning}

\author{Qiushi Zhu, Xiaoying Zhao, Jie Zhang, Yu Gu, Chao Weng, Yuchen Hu
\thanks{Manuscript created May 2023; This work is supported by the National Natural Science Foundation of China (62101523), Hefei Municipal Natural Science Foundation (2022012), USTC Research Funds of the Double First-Class Initiative (YD2100002008). (Corresponding author: {\it Jie Zhang})}
\thanks{Qiushi Zhu, Xiaoying Zhao and Jie Zhang are with the Department of Electronic Engineering and Information Science, University of Science and Technology of China (USTC), Hefei 230026, China (e-mail: \{qszhu, xyzhao1123\}@mail.ustc.edu.cn;
jzhang6@ustc.edu.cn). 
Yu Gu and Chao Weng are with Tencent AI LAB (e-mail:\{colinygu, cweng\}@tencent.com).
Yuchen Hu is  with the School of Computer Science and Engineering, Nanyang Technological University (NTU), Singapore 639798 (e-mail: yuchen005@e.ntu.edu.sg).}
}

\markboth{Journal of \LaTeX\ Class Files, Vol. xx, No. xx, May 2023}
{Shell \MakeLowercase{\textit{et al.}}: Bare Demo of IEEEtran.cls for IEEE Journals}
\maketitle

\begin{abstract}
Recently, many efforts have been made to explore how the brain processes speech using electroencephalographic (EEG) signals, where deep learning-based approaches were shown to be applicable in this field.
In order to decode speech signals from EEG signals, linear networks, convolutional neural networks (CNN) and long short-term memory networks are often used in a supervised manner. Recording EEG-speech labeled data is rather time-consuming and laborious, while unlabeled EEG data is abundantly available. Whether self-supervised methods are helpful to learn EEG representation to boost the performance of EEG auditory-related tasks has not been well explored.
In this work, we first propose a self-supervised model based on contrastive loss  and reconstruction loss to learn EEG representations, and then use the obtained pre-trained model as a feature extractor for downstream tasks.
Second, for two considered downstream tasks, we use CNNs and Transformer networks to learn local features and global features, respectively.
Finally, the EEG data from other channels are mixed into the chosen EEG data for augmentation.
The effectiveness of our method is veriﬁed on the EEG match-mismatch and EEG regression tasks of the ICASSP2023 Auditory EEG Challenge. 

\end{abstract}


\IEEEpeerreviewmaketitle

\section{Introduction}
\label{sec:intro}
\IEEEPARstart{I}{n human} auditory systems, ions flowing through neuronal cells in the brain form electrical currents that generate electroencephalographic (EEG) activity.
EEG signals are commonly-used in medical fields, e.g., Alzheimer's diagnosis~\cite{jeong2004eeg}, sleep detection~\cite{zhu2014analysis}, emotion recognition~\cite{jenke2014feature}, speech enhancement~\cite{zhang2023basen}.
In the field of speech, recent studies have concentrated on detecting neural tracking of speech features in the EEG, in order to understand how the brain processes speech~\cite{27965557,DILIBERTO20152457,vanthornhout2018speech,puffay2023relating,9743715,9467380}.
Neural tracking has been used for a variety of acoustic representations of speech, e.g., spectrograms~\cite{DILIBERTO20152457,jalilpour2021extracting} or envelope representation~\cite{vanthornhout2018speech,accou2023decoding,9287417,Accou_2021}.
The relationship between speech envelopes and speech understanding was revealed in~\cite{vanthornhout2018speech,Accou_2021,iotzov2019eeg}.
In addition, various models were proposed to reconstruct speech (i.e., stimulus signal) from EEG signals~\cite{27965557,vanthornhout2018speech,accou2023decoding,thornton2022robust} (called forward regression), reconstruct EEG signals from speech~\cite{lesenfants2019predicting} (called backward regression), and distinguish whether EEG signals match speech signals~\cite{9287417,Accou_2021,de2018decoding,de2021auditory,jalilpour2021extracting} (called classification).

For the forward estimation task, it is often-used to present the natural speech to the listener and simultaneously record the listener's EEG signals, and then decode the information related to the speech signal from the listener's EEG signals.
Linear models~\cite{vanthornhout2018speech} can be leveraged to reconstruct speech envelope representations from EEG signals. It was shown that using natural speech as a stimulus for electrophysiological measures of hearing is a paradigm shift, which can bridge the gap between behavioral and electrophysiological measures.
However, linear models are difficult to handle complex nonlinear relationships, so many neural network-based models were thus proposed, which significantly outperform the former in different auditory scenarios~\cite{thornton2022robust}.
A typical example is the convolutional neural networks (CNNs) based  very large augmented auditory inference (VLAAI) model that was proposed in~\cite{accou2023decoding}.
For the EEG match-mismatch task, a long short-term memory (LSTM) based model was proposed in~\cite{jalilpour2021extracting}, where the impact of level-wise speech features (e.g., envelope, mel spectrogram, voice activity, phoneme identity) on the performance was analyzed.
In~\cite{9287417}, a dilated convolution-based model was proposed to maximize the receptive field for modeling complex nonlinear relationships, leading to an improvement over both the linear and convolutional models.
In~\cite{Accou_2021}, it was validated that the model with dilated convolution outperforms the linear model.

Self-supervised methods are able to utilize large amounts of unlabeled data and  are effective for many tasks in the fields of both speech~\cite{NEURIPS2020_92d1e1eb,9585401,baevski2022data2vec} and EEG~\cite{kostas2021bendr,chien2022maeeg,mohsenvand2020contrastive,NEURIPS2022_d81ecfc8}.
Typical self-supervised speech representation learning methods include Wav2vec2.0~\cite{NEURIPS2020_92d1e1eb}, HuBERT~\cite{9585401} and data2vec~\cite{baevski2022data2vec}.
There are also some models based on self-supervised methods for EEG signal processing. For example, inspired by wav2vec2.0, BENDR proposed in~\cite{kostas2021bendr} uses a contrastive learning approach and outperforms previous methods on the sleep stage classification task. The self-supervised MAEEG model in~\cite{chien2022maeeg} uses the Transformer network to significantly improve the accuracy of sleep classification by reconstructing the masked EEG features.
In~\cite{mohsenvand2020contrastive}, a contrastive EEG representation learning framework was proposed, which augments the samples by recombining data from different channels, showing applicability in e.g., emotion recognition, normal/non-normal EEG classification and sleep stage classification.

Clearly, there is a lack in the exploration of  self-supervised methods for EEG match-mismatch and EEG regression tasks.
Inspired by~\cite{NEURIPS2020_92d1e1eb}, we first propose contrastive learning-based and reconstruction-based Eeg2vec models to learn EEG representations, and then use the obtained pre-trained model as a feature extractor for EEG match-mismatch  and EEG regression tasks.
Second, for the backbone models of the two tasks, we use multiple convolutional and self-attention modules to learn local and global features, respectively, which are able to acquire information at different granularities and exceed the performance of the dilated convolution-based models.
Finally, adding a certain percentage of EEG data from other channels to the EEG data can effectively augment the data diversity and improve the model performance.
The effectiveness of our method is demonstrated on both match-mismatch  and regression tasks of the ICASSP2023 Auditory EEG Challenge\footnote{https://exporl.github.io/auditory-eeg-challenge-2023/}.

\begin{figure}[!t]
	\centering
	\includegraphics[width=0.45\textwidth]{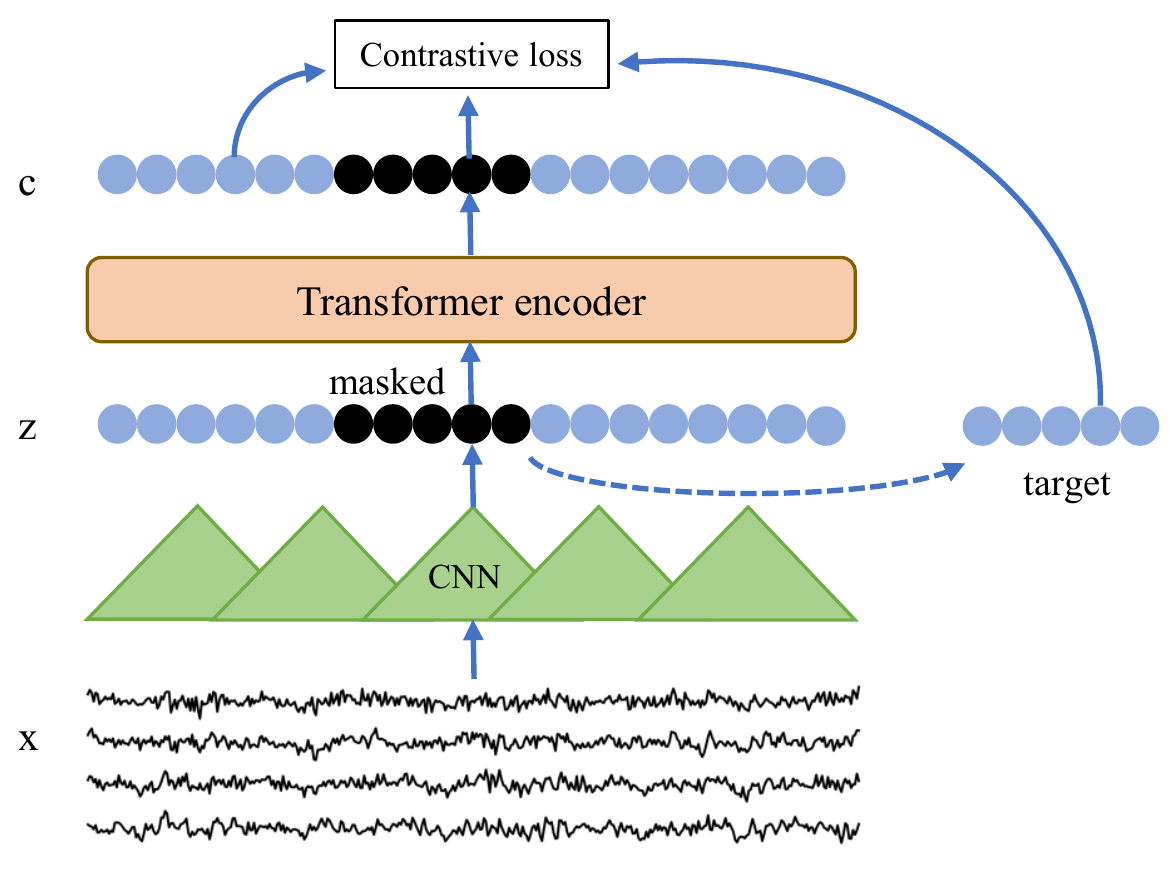}
	\caption{The structure of the self-supervised pre-training Eeg2vec model.}
	\label{fig:Eeg2vec}
\end{figure}

\section{Methodology}

\subsection{Self-supervised Eeg2vec pre-training model}

The proposed Eeg2vec model is shown in Fig.~\ref{fig:Eeg2vec}, including the feature encoder and context modules.
The feature encoder module consists of CNN layers, layer normalization, GELU activation function, and the context module mainly consists of a Transformer encoder.
Specifically, the 64-channel EEG signal $x$ is fed to the CNN layer to obtain the EEG feature $z$. The masked feature $z$ is then fed to the Transformer encoder to learn the context information $c$. The true feature $z_{tar}$ at the masked position is taken as the target for self-supervised learning.
The goal of contrastive learning is to expect the predicted information to be as close as possible to the positive samples and as far as possible from the negative samples.
For the masked time step $t$, given the predicted context representation $c_{{\rm pre}_t}$, the target $c_{{\rm tar}_t}$ and the set of negative samples $c_{ns}$, the contrastive loss $\mathcal{L}_{c}$ is given by

\begin{equation}
	\mathcal{L}_{c} = -\log \frac{\exp({\rm sim}(c_{{\rm pre}_t},z_{{\rm tar}_{t}})/\kappa)}{\sum_{\tilde{c} {\sim} {\{z_{\rm tar}, c_{ns}\}}}\exp({\rm sim}(c_{{\rm pre}_t},\tilde{c})/\kappa)},
	\label{eq1}
\end{equation}
where $c_{ns}$ is obtained by random sampling from the contextual representation $c$ at other positions,
$\rm sim$ denotes the cosine similarity and $\kappa$ is the temperature coefficient. For more details on the contrastive loss, please refer to~\cite{NEURIPS2020_92d1e1eb}.
In addition, we also consider EEG representation learning as a reconstruction task, where the contrastive loss function in the Eeg2vec model is replaced with a reconstruction loss function, given by
\begin{equation}
	\mathcal{L}_{r} = \frac{1}{T}\sum_{t=1}^{T} |c_{{\rm pre}_t} - z_{{\rm tar}_t} |,
\end{equation}
where $T$ denotes the number of all-time steps. These two loss functions will be compared experimentally.

\begin{figure}[!t]
	\centering
	\includegraphics[width=0.45\textwidth]{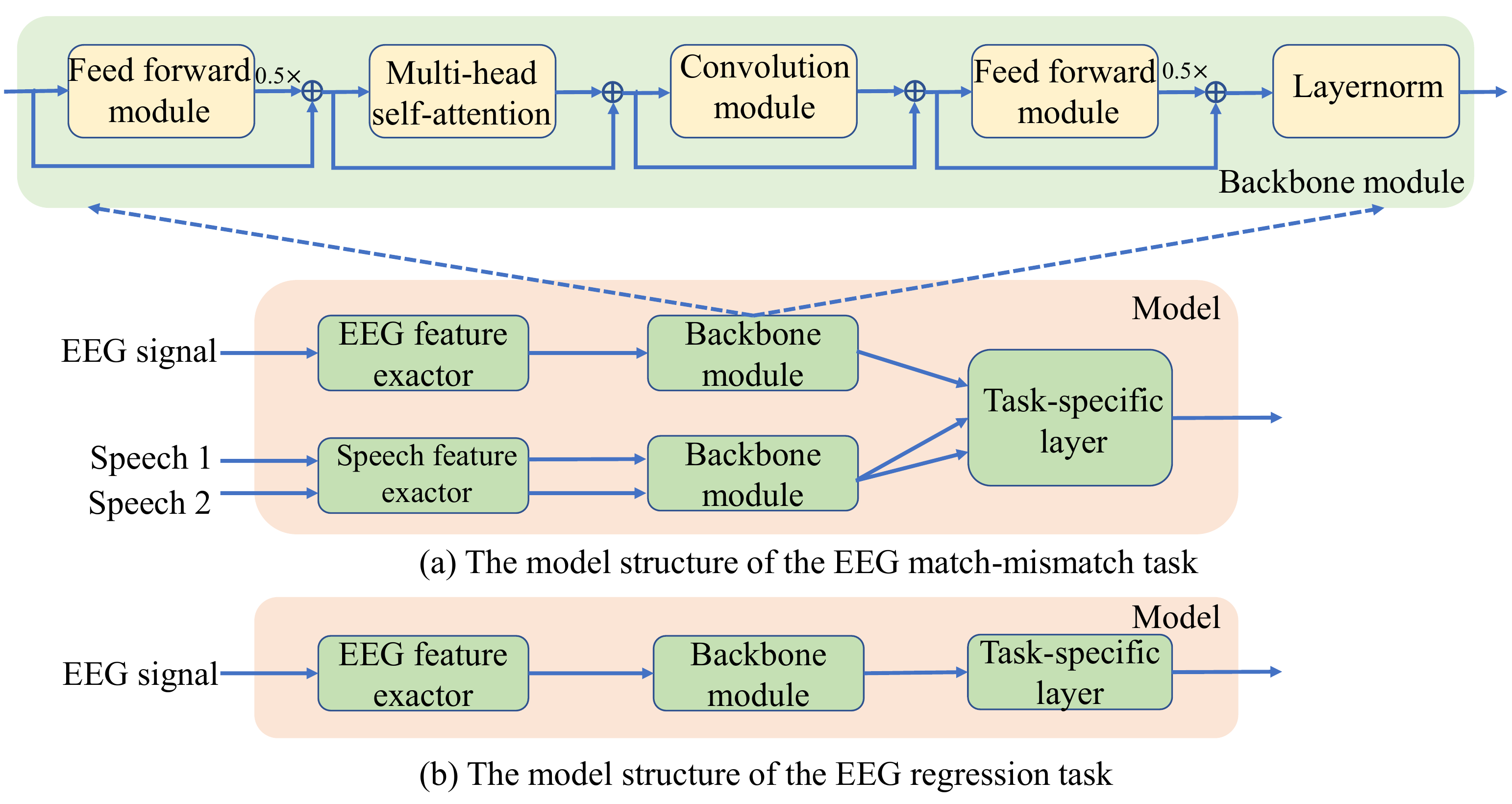}
	\caption{The model structures for (a) the EEG match-mismatch task and (b) the EEG regression task.}
	\label{fig:taskmodel}
\end{figure}
	
\subsection{Model structures of two downstream tasks}
After pre-training, we take the pre-trained model as a feature extractor with frozen parameters for the EEG signals.
To better model the local and global information contained in EEG signals, we use CNN and self-attention modules.
For the match-mismatch and regression tasks, the model structures are shown in 
Fig.~\ref{fig:taskmodel}(a) and Fig.~\ref{fig:taskmodel}(b), respectively.	
The Backbone module is a multi-layer Conformer encoder, which mainly contains two half-step feed forward blocks, a self-attention block and a convolution block.	
For more details on the Conformer encoder, please refer to~\cite{gulati20_interspeech}.
For the task-specific layer of the EEG match-mismatch task, the cosine similarity is computed separately for the EEG features and the two speech features, which are then concatenated and fed into the linear projection layer.
The task-specific layer of the EEG regression task mainly contains linear projection layers.

\begin{figure}[!t]
	\centering
	\includegraphics[width=0.45\textwidth]{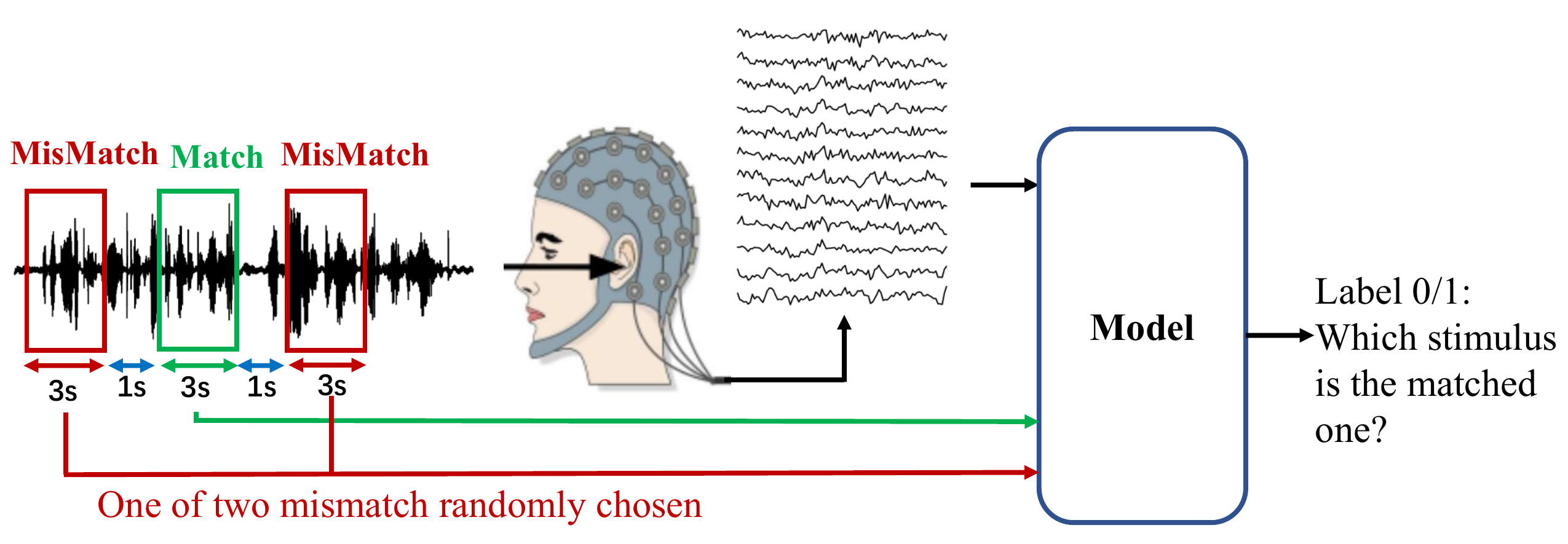}
	\caption{The system diagram for the match-mismatched classification task.}
	\label{fig:matchmismatch}
\end{figure}

\textbf{\emph{EEG Match-mismatch Task:}}
This is indeed a binary classification task, where the objective is to determine which input stimulus segment corresponds to the EEG signal.
As shown in Fig.~\ref{fig:matchmismatch}, the model has three inputs: a segment of EEG signal, a segment of aligned speech stimuli and a segment of imposter speech stimuli.
The model  distinguishes which stimulus segment corresponds to the EEG signal based on the usage of the typical binary cross entropy for optimization.

\begin{figure}[!t]
	\centering
	\includegraphics[width=0.45\textwidth]{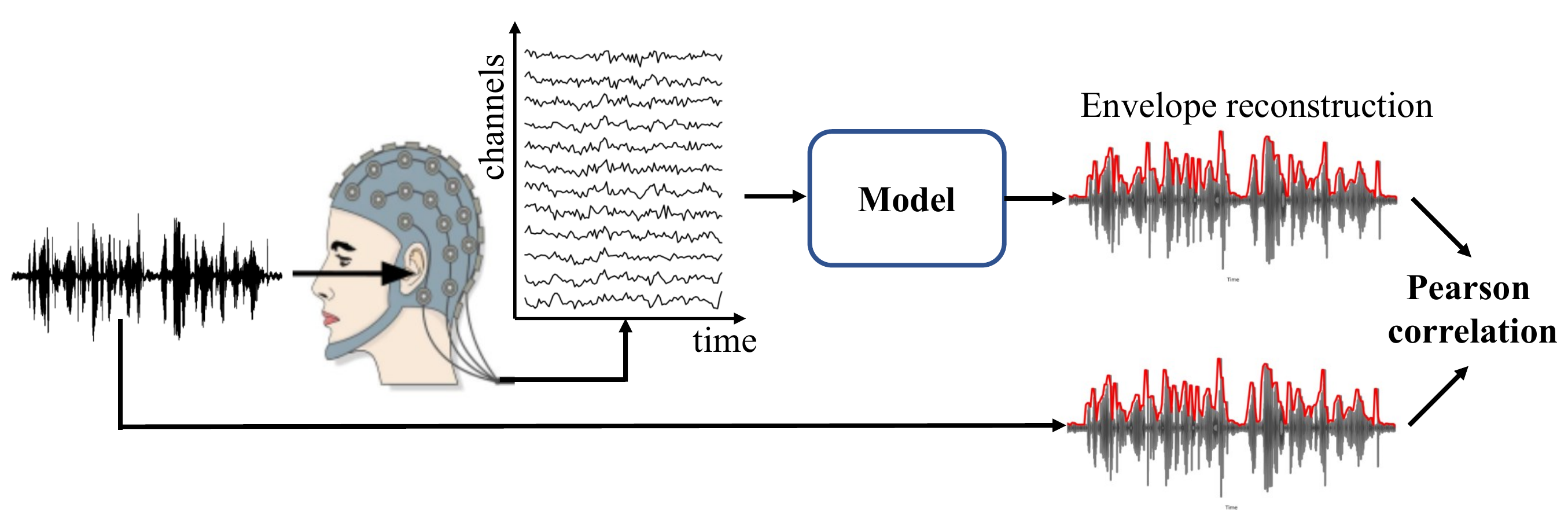}
	\caption{The diagram of speech envelope reconstruction using EEG signals.}
	\label{fig:regression}
\end{figure}

\textbf{\emph{EEG Regression Task:}}
For the regression task, the objective is to reconstruct the stimulus signal from the EEG signal.
As shown in Fig.~\ref{fig:regression}, a speech stimulus signal $y$ is sent to the human ear, and its corresponding EEG signals $x$ are collected, which are fed into the model to reconstruct the speech envelope $\hat{y}$.
The considered regression criterion is the Pearson correlation coefficient (PCC), given by
\begin{equation}
	\rho_{y,\hat{y}}=\frac{{\rm cov}(y,\hat{y})}{\sigma_{y}\sigma_{\hat{y}}},
	\label{eq3}
\end{equation}
where $\rm cov$  denotes the covariance and $\sigma$  the standard variance.
The larger the PCC between the reconstructed and real speech envelopes, the better the regression performance.
The loss function for this task is the negative PCC.

\section{Experimental Setup}

\textbf{\emph{Dataset:}}
The dataset used in experiments is available online\footnote{https://exporl.github.io/auditory-eeg-challenge-2023/dataset/}, where the EEG data were measured in a well-controlled laboratory environment (soundproof and electromagnetically shielded room) using a high-quality 64-channel Biosemi ActiveTwo EEG recording system with 64 active Ag-AgCl electrodes and two additional electrodes. All 64 electrodes were placed according to international standards.
The dataset was collected from 85 young normal-hearing subjects (all hearing thresholds $\leq$ 25 dB HI) whose native language was Dutch. Subjects with any neurological or hearing-related medical history were excluded from the study.
Each participant listened to 8 to 10 trials, with each trial lasting approximately 15 minutes. The order of the trials was randomly assigned among the participants.
The stimuli were podcasts or audiobooks. All stimuli were single-person stories told in Flemish (Belgian Dutch) by a native Flemish speaker. The stimuli were varied across subjects to obtain a wide range of speech material. 

The training set includes a total of 508 trials from 71 subjects, using 57 different stimuli. The total amount is 7216 minutes (120 hours). The training set is shared between the two tasks.
The test set consists of two parts: held-out stories and held-out subjects, which are split into two separate parts. This ensures that the test sets for the two tasks do not overlap.
Test Set 1 (held-out stories) consists of data for the 71 subjects (sub-01 to sub-71) who are included in the training set. For each group of subjects, they retain one story that does not appear in the training set, resulting in a total of 944 minutes.
Test Set 2 (held-out subjects) comprises data for 14 subjects (sub-72 to sub-85) who are not included in the training set and hence referred to as held-out subjects. The data for these subjects were collected using the same protocol as for the other 71 subjects and amounted to a total of 1260 minutes.

\textbf{\emph{Data Preprocessing:}}	
The EEG signal was downsampled from 8192 Hz to 1024 Hz and artifacts were removed using a multichannel Wiener filter. The signal was then re-referenced to a common average and downsampled further to 64 Hz. These standard steps are frequently employed in EEG signal processing.	
For the regression task, this dataset includes a specific version of the envelope, which is thus used to evaluate the performance directly. To estimate the envelope, we utilize a gammatone filter bank consisting of 28 subbands with an equivalent bandwidth spacing and center frequencies ranging from 50 Hz to 5 kHz. Subsequently, we take the absolute value of each sample in the filters and raise them to the power of 0.6. The resulting values for all subbands are then averaged to obtain a single speech envelope. Finally, we downsample the resulting envelope to 64 Hz.

\textbf{\emph{Training details:}}
For Eeg2vec, the feature encoder contains 4 convolutional layers with a  kernel size of 3 and a  dimension of 64, and the context module contains 12 Transformer encoder layers, where the self-attention module has a dimension of 256 and the feed-forward module has a dimension of 1024.
The ratio of masking on the feature is set to 0.5 and the length of masking is set to 10.
In (\ref{eq1}), the number of negative samples is 50 and the temperature coefficient $\kappa$ is set to 1.5.
The data used for pre-training is the entire training set, the number of updates for pre-training is 100k, the Adam~\cite{kingma2014adam} optimizer is used, and the learning rate is set to 5e-4.
For the two tasks, the EEG feature extractor is the pre-trained Eeg2vec model, and the speech feature extractor is two 1-dimensional convolutional layers with a kernel size of 3 and a dimension of 64.
For the backbone models, the dimensions of the self-attention and feed-forward modules are 128 and 512, respectively, the convolution kernel size is 31, and the dimension of the convolution is 128.
The models are trained for 100 epochs with the AdamW~\cite{loshchilov2018decoupled} optimizer with a learning rate of 1e-4.

\textbf{\emph{Evaluation metrics:}} We use the official evaluation metrics of ICASSP2023 Auditory EEG Challenge. Specifically, for the \textbf{match-mismatch} task, we  first compute the average accuracy score for each subject as
\begin{equation}
ACC_s = \sum_{i=0}^{n_s}[{\rm label}_\text{predicted} = {\rm label}_\text{true}]/n_s.
\end{equation}
The average scores on the two test sets are then $S_1 = \sum_{s=1}^{71}\frac{ACC_s}{71}$ and $S_2 = \sum_{s=72}^{85}\frac{ACC_s}{14}$, respectively. 
The final score is defined as
\begin{equation}
	A_{score} = \frac{2}{3}S_1 + \frac{1}{3}S_2.
\end{equation}

For the \textbf{regression} task, we first compute the PCC scores for all subjects, which are then averaged over two test sets  similarly to the calculation of ACC. The two set-dependent scores are finally weighted summed as $P_{score} = \frac{2}{3}P_1 + \frac{1}{3}P_2$.

\begin{table}[]
\caption{The experimental results on EEG match-mismatch and regression tasks.}
\label{tab:table1}
\centering
\begin{tabular}{l|cc|cc}
\hline
\multirow{3}{*}{\textbf{Method}} & \multicolumn{2}{c|}{\textbf{Match-mismatch task}} & \multicolumn{2}{c}{\textbf{Regression task}}      \\ \cline{2-5} 
                                 & \multicolumn{2}{c|}{$\textbf{A}_{score}$}            & \multicolumn{2}{c}{$\textbf{P}_{score}$}             \\ \cline{2-5} 
                                 & \multicolumn{1}{c|}{\textbf{dev}} & \textbf{test} & \multicolumn{1}{c|}{\textbf{dev}} & \textbf{test} \\ \hline
Rank1                            & \multicolumn{1}{c|}{-}            & 82.13         & \multicolumn{1}{c|}{-}            & 0.1589        \\
Rank2                            & \multicolumn{1}{c|}{-}            & 79.05         & \multicolumn{1}{c|}{-}            & 0.1535        \\
Rank3                            & \multicolumn{1}{c|}{-}            & 78.94         & \multicolumn{1}{c|}{-}            & 0.1519        \\ \hline
Official baseline~\cite{9287417}                & \multicolumn{1}{c|}{-}            & 77.51         & \multicolumn{1}{c|}{-}            & 0.1023        \\
Our baseline                     & \multicolumn{1}{c|}{76.03}        & 77.25         & \multicolumn{1}{c|}{0.1951}       & 0.1446        \\
Black\_box (Our)        & \multicolumn{1}{c|}{76.82}        & 78.69         & \multicolumn{1}{c|}{0.2011}       & 0.1519        \\
Our final                        & \multicolumn{1}{c|}{78.94}        & 79.76         & \multicolumn{1}{c|}{0.2127}       & 0.1638        \\ \hline
\end{tabular}
\end{table}

\section{Experimental results}

\textbf{Main Results:} The results of our model and the baseline model for the match-mismatch and regression tasks are shown in Table~\ref{tab:table1}. 
For the convenience of comparison, we also include the results of the top three teams in the challenge, which are named Rank1, Rank2 and Rank3 in order.
For the match-mismatch task, the official baseline model utilizes a dilated convolutional network-based  model, which achieves an $A_{score}$ of 77.51 on the test set, while no results are reported on the validation set.
Our baseline model is based on the Conformer structure and does not use a pre-trained model as a feature extractor, which is able to achieve an $A_{score}$ of 77.25 on the test set.
The performance of our baseline model is slightly lower than that of the official baseline, mainly  because of insufficient tuning of model parameters and the number of submission limits.
The results (\textbf{Black\_box}) that we submitted to the leaderboard\footnote{https://exporl.github.io/auditory-eeg-challenge-2023/task1/leaderboard/} show that $A_{score}$ reaches 78.69 on the test set, the fourth place in the ranking, with the Eeg2vec model depending on the reconstruction loss as the feature extractor.
Our final model, which utilizes the contrastive learning-based Eeg2vec with data augmentation and more unlabeled data, can further increase the accuracy to 78.94 on the validation set and 79.76 on the test set, respectively.

For the regression task, the official baseline uses a linear model, which achieves a $P_{score}$ of 0.1023 on the test set, while no results were reported on the validation set.
Our baseline model is based on the Conformer and does not use a pre-trained model as the feature extractor, which outperforms the official baseline model.
The results (\textbf{Black\_box}) show that $P_{score}$  reaches 0.1519 on the test set, obtaining the third place in the leaderboard\footnote{https://exporl.github.io/auditory-eeg-challenge-2023/task2/leaderboard/}. The model also uses the Eeg2vec model based on the reconstruction method as the feature extractor.
Our final model obtains the $P_{score}$ of 0.2127 on the validation set and 0.1638 on the test set  with the Eeg2vec model based on contrastive learning and data augmentation techniques.

\textbf{Ablation Study:}
In order to more clearly understand the impact of different configurations on the results, we conduct comparison experiments on the match-mismatch task  in Table~\ref{tab:table2}.
When the reconstruction-based pre-trained Eeg2vec model is used as the feature extractor, the $A_{score}$ on the validation set can reach 76.82. In case  the contrastive learning-based pre-trained Eeg2vec model  is used as the feature extractor, the classification accuracy (e.g., 77.25) becomes higher. This indicates that the contrastive learning method can learn more informative contextual representations.
Based on the pre-trained Eeg2vec model, mixing EEG data from different channels to augment the dataset shows a further match-mismatch performance improvement.

\begin{table}[]
\caption{The $\mathbf{A}_{score}$  result of our model on match-mismatch task with different configurations.}
\label{tab:table2}
\centering
\resizebox{0.98\columnwidth}{!}{
\begin{tabular}{l|c|c|cl}
\hline
{\textbf{Method}}   &EEG length (s)  & More data & \textbf{dev} & {\textbf{test}}                                                                                        \\ \hline
Our baseline                                                                                        & 3                                                                                        & -                                                                                        & 76.03        & 77.25                             \\ \cline{1-1}
Reconst. Eeg2vec                            & 3                                                                                        & -                                                                                        & 76.82        & 78.69                             \\ \cline{1-1}
Contrast.  Eeg2vec                                & 3                                                                                        & -                                                                                        & 77.25        & 79.05                             \\ \hline
\multirow{4}{*}{\begin{tabular}[c]{@{}l@{}}Contrast. Eeg2vec\\ + Data augment\end{tabular}} & 3                                                                                        & -                                                                                        & 77.69        & 79.30                             \\
                                                                                                    & 4                                                                                        & -                                                                                        & 78.82        & 79.62                             \\
                                                                                                    & 5                                                                                        & -                                                                                        & 79.41        & 80.01                             \\ \cline{2-5} 
                                                                                                    & 3                                                                                        & \checkmark                                                                                & 78.94        & 79.76                             \\ \hline
\end{tabular}}
\end{table}

The length of the input EEG segment used in the challenge was fixed to 3 seconds for the match-mismatch task. To observe the effect of the EEG segment length on the results, we compare the EEG input lengths on the performance in Table~\ref{tab:table2}.
We find that the model can perform better with an increase in the EEG segment length, due to more information contained in a longer EEG segment.
This also indicates that choosing the right EEG segment is crucial for the classification task, which is consistent with the conclusion in~\cite{Accou_2021}.

In addition to this dataset, we utilize more unlabeled EEG data from~\cite{BRODERICK2018803} for self-supervised pre-training, which is processed similarly as this dataset, and the  result is shown at the bottom of Table~\ref{tab:table2}.
It is clear that using more unlabeled data can significantly improve the quality of EEG representation with the same EEG input length (e.g., $A_{score}$ 77.69 vs 78.94 on the dev set). The resulting $A_{score}$ of using more unlabeled data becomes 79.76 on the test set of the match-mismatch task (see Table I).
Although many participants have explored different methods in the challenge, the improvement in the Pearson coefficient for the EEG regression task is rather marginal, and we will continue to explore this task in the future.
Although self-supervised pre-training methods have shown powerful capabilities in other fields, further research is needed in the field of EEG signal processing, such as how to learn better  EEG representations, which may require more specialized EEG knowledge.

\section{CONCLUSION}
\label{sec:conclusion}
In this paper, we proposed self-supervised models based on contrastive learning and reconstruction loss to learn EEG representations, which were applied as front-end feature extractors to the EEG match-mismatch and  regression tasks.
Both self-supervised models were effective in boosting the performance of downstream tasks, while the representation learned by the contrastive learning based model was shown to be more informative. We found that
using CNN and Transformer networks can effectively learn local and global features in EEG signals.
In addition, the mixing of EEG data from different channels can effectively augment the dataset and improve the model performance. As in this work we followed the typical Wiener filter to pre-process the recorded EEG signals that might contain various random measurement noises, in the future we will investigate artefacts removal methods with an application to these EEG-auditory tasks.

\bibliographystyle{IEEEbib}
\bibliography{strings,refs}
\end{document}